\documentclass[aps,prl,showpacs,amsmath,amsfonts,superscriptaddress,twocolumn]{revtex4}
\usepackage{graphicx}
\usepackage{dcolumn}
\usepackage{bm,color}
\usepackage{pslatex}
\usepackage{hyperref}
\begin{document}

\title{Diagrammatic Monte Carlo for Correlated Fermions}

\author{E.~Kozik}
\affiliation{Theoretische Physik, ETH Zurich, 8093 Zurich, Switzerland }
\author{K.~Van Houcke}
\email{kris.vanhoucke@ugent.be}
\affiliation{Department of Physics, University of Massachusetts, Amherst, MA 01003, USA}
\affiliation{Universiteit Gent - UGent, Vakgroep Subatomaire en
  Stralingsfysica \\
  Proeftuinstraat 86, B-9000 Gent, Belgium}
\author{E. Gull}
\affiliation{Department of Physics, Columbia University, New York, NY 10027, USA}
\author{L. Pollet}
\affiliation{Department of Physics, University of Massachusetts, Amherst, MA 01003, USA}
\affiliation{Physics Department, Harvard University, Cambridge, Massachusetts 02138, USA}
\author{N. Prokof'ev}
\author{B. Svistunov}
\affiliation{Department of Physics, University of Massachusetts, Amherst, MA 01003, USA}
\affiliation{Russian Research Center  ÒKurchatov InstituteÓ, 123182 Moscow, Russia }
\author{M. Troyer}
\affiliation{Theoretische Physik, ETH Z\"urich, 8093 Z\"urich, Switzerland }

\date{\today}

\begin{abstract}
We show that Monte Carlo sampling of the Feynman diagrammatic series (DiagMC) can be used for tackling
hard fermionic quantum many-body problems in the thermodynamic limit by presenting accurate results for the repulsive Hubbard model in the correlated Fermi liquid regime.  Sampling Feynman's diagrammatic series for the single-particle self-energy  we can study moderate values of the on-site repulsion ($U/t \sim 4$) and temperatures down to $T/t=1/40$. We compare our results with high temperature series expansion and with single-site and cluster dynamical mean-field theory.
\end{abstract}



\maketitle

Advancing first-principle simulations of fermionic many-particle systems is notoriously difficult due to an exponential computational complexity \cite{Troyer}. One of the most famous examples is the Hubbard model~\cite{Hubbard63, Anderson} whose phase diagram remains highly controversial,
\begin{equation}
\hat{H} = -t \sum_{\langle i,j\rangle, \sigma}    \hat{c}_{i \sigma}^{\dagger} \hat{c}_{j \sigma}^{\phantom{\dagger}}
  + U \sum_i  \hat{n}_{i \uparrow} \hat{n}_{i \downarrow} - \mu \sum_{i, \sigma} n_{i \sigma}\; ,
\label{FH}
\end{equation}
where $\hat{c}_{i \sigma}^{\dagger}$
creates
a fermion with spin projection $\sigma=\uparrow,  \downarrow$ on site $i$,
$ \hat{n}_{i \sigma} = \hat{c}_{i \sigma}^{\dagger}\hat{c}_{i \sigma}^{\phantom{\dagger}}$,
$\langle \dots \rangle$ restricts summation to neighboring lattice sites, and $t$, $U$ and $\mu$ are the hopping amplitude, the on-site repulsion, and the chemical potential respectively.

The Hubbard model is of great technological and scientific importance since it is believed by many to be
the central model for high-temperature superconductors \cite{Anderson}. Moreover, it can nowadays be
realized with ultracold atoms in optical lattices \cite{Esslinger05} where Mott physics~\cite{Jordens08, Schneider08} has been observed recently. The prospect of using cold atomic gases as building blocks of
quantum simulators is thus becoming a reality. However, claiming the equation of state, either numerically or experimentally, can only be substantiated on the basis of unbiased tools. Not only numerical methods need to be tested against known unbiased results, but also quantum emulators need to be validated by
benchmarking them at ever lower temperatures. In the bosonic case this can be done with high precision~\cite{Trotzky09}, but for fermions it is still an open question what method can be used
to accurately describe the low-temperature regime of the model. Precise numerical solutions of the Hubbard model can only be obtained in the particle-hole symmetric case $\langle n_\sigma \rangle = 1/2$, where the sign problem is absent in quantum Monte Carlo (QMC) simulations. In all other cases circumventing exponential scaling necessarily involves approximations with systematic errors that are hard to assess and control.

In general terms, Diagrammatic Monte Carlo (DiagMC) \cite{Prokofev98} is a numerical approach based on stochastic sampling of Feynman diagrams \cite{Fetter, Landau} to increasingly higher orders in the coupling constant---rather than evaluating integrals over internal variables
for each diagram, one samples their momenta, imaginary times and expansion orders stochastically. So far, DiagMC algorithms have been successfully used for obtaining unbiased solutions for polaron problems (for the latest work, see Ref.~\cite{Prokofev08}).

Nevertheless, the crucial question remained unclear of whether the DiagMC approach could be applied to true \textit{many-body} systems in the \textit{thermodynamic limit} at physically interesting temperatures much smaller than the Fermi energy. The convergence of a diagrammatic series for the many-body system is not guaranteed---in many cases the series has zero convergence radius, which can be revealed by Dyson's system collapse argument  \cite{Dyson}.
Fortunately, the argument does not apply to the Hubbard model. However, even then the convergence is not guaranteed \textit{a priori} due to possible \textit{unphysical} singularities in the parameter space, as, e.g., is the case with HTSE \cite{Henderson92, tenHaaf92}, which for the Hubbard model diverges at $T \sim t$.

An algorithm for the many-body problem, based on a DiagMC sampling of the self-energy, was developed recently by some of us \cite{VanHoucke08}. In this Letter, we find that this algorithm can produce controlled and accurate results (with reliable error bars) in the correlated Fermi liquid regime of the Hubbard model at moderate interactions in any spatial dimension.
Comparison with single-site \cite{Georges96} and cluster \cite{Hettler98,Lichtenstein00,Kotliar01,Maier05}
dynamical mean-field theory in the dynamical cluster approximation (DCA) variant in 2D and with
high-temperature series expansions (HTSE) \cite{Henderson92, tenHaaf92} reveals an impressive agreement,
thus proving theoretical control over the Fermi liquid regime of the model at moderate values of $U/t$.
Our simulations demonstrate that the diagrammatic expansion for the Hubbard model does converge (at least for moderate interactions) down to temperatures much smaller than the Fermi energy \footnote{In a more general case, with the internal dressing of propagators and interaction lines (skeleton series), explicit symmetry-breaking terms, and resummation techniques (which were shown recently to be perfectly compatible with DiagMC \cite{Prokofev08,Prokofev07}),  physical answers can in principle be extracted from diagrammatic series even outside of their finite convergence radius.}.

Despite the inherent sign problem, our scheme is capable of reaching diagram orders sufficient to claim the convergence in the thermodynamic limit, the feasibility of which for a many-body fermionic system was not clear \textit{a priori}. The ability to calculate higher-order diagrams is also important, e.g., in describing decay processes in QCD (see, e.g. \cite{SCET}), and might prove relevant for solving Dyson-Schwinger equations in QED and QCD \cite{D-Sch}, where there is a growing interest in numerical perturbative expansions \cite{NPT}.

\begin{figure}[t!]
\begin{center}
\includegraphics[angle=0, width=0.9\columnwidth]{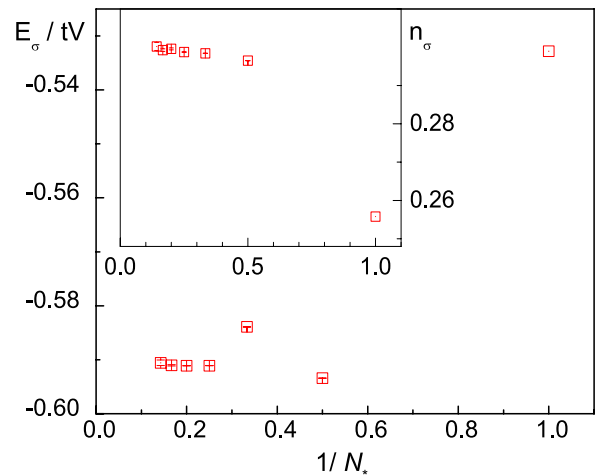}
\caption{ \label{fig:2d_conv} (color online) Energy density per spin component for the 2D Hubbard model as a function of inverse expansion order cut-off $N_*$. This example is showing data for  $U/t=4$ and $T/t= 0.025$.
The density is shown in the inset. We observe convergence around order $4$,
and the extrapolation to the infinite order can be done reliably. Error bars are smaller than the symbol size.
}
\end{center}
\end{figure}

The crucial difference between the thermodynamic-limit DiagMC and other Monte Carlo approaches---dealing with a finite-size cluster of
volume $V$ in the $(d+1)$-dimensional space---is that in DiagMC the complexity of the sign problem coming from sign-alternating diagrammatic
expressions is linked to the {\it expansion order} of the Feynman perturbation series and not to the large volume $V$ required for the thermodynamic limit extrapolation.
Away from criticality and for moderate interaction strength, only a few diagram orders are needed until convergence of the diagrammatic series
is reached (see Fig.~\ref{fig:2d_conv}). This greatly reduces the computational cost, the major part of which is nothing but ensuring that higher
order diagrams cancel each other and thus do not contribute to the answer.

To demonstrate the power of DiagMC we will compare it to other
state-of-the-art first-principle approaches:
high-temperature series expansion
and dynamical mean field theory (DMFT) in the single-site and dynamic cluster
approximation (DCA) implementation \cite{Georges96, Maier05}.
DMFT and its cluster extensions have so far arguably been the method of choice for studies of the Hubbard model, producing a phase diagram in close agreement with that of the cuprates~\cite{Maier05}. Single-site DMFT  treats the correlated many-body system as a quantum impurity problem where each site can be viewed as an impurity embedded self-consistently in a reservoir produced by the other sites. It goes beyond conventional mean-field theory by taking all temporal correlations into account, but neglects the momentum dependence of the self-energy $\Sigma$.
Short-range correlations can be taken into account by extending the size of the impurity, leading to various cluster schemes.

HTSE is a method where the partition function $Z = {\rm Tr}\, \exp(-\beta \hat{H})$ is expanded in $\beta t$ around the atomic limit ($t=0$) and all the terms are summed up exactly up to a maximum expansion order. This expansion is analytic and yields therefore precise thermodynamics of the Hubbard model, provided the series converges at the maximum expansion order considered.

We consider the two cases of moderately $(20\%)$ and heavily (40\%)
doped regimes and moderate values of the interaction strength ($U/t=4$) in two and three dimensions.
The convergence of DiagMC is checked by performing simulations with ever increasing expansion order
cut-off until the extrapolation to infinite order can be done with confidence.
Although the sign problem is manageable for moderate interactions, simulations do get exponentially
harder when the expansion order is increased. Fortunately, we can extract the converged
result before the sign problem becomes too severe. In case of the Hubbard model we typically see convergence of the series around fourth order for a moderate interaction $U/t=4$ and
sufficiently far away from half filling (see Fig.~\ref{fig:2d_conv}).
All the results shown in this work are obtained by observing a clear plateau, from which we deduce the
infinite-order answer with confidence.

\begin{figure}[t!]
\begin{center}
\includegraphics[angle=0, width=0.95\columnwidth]{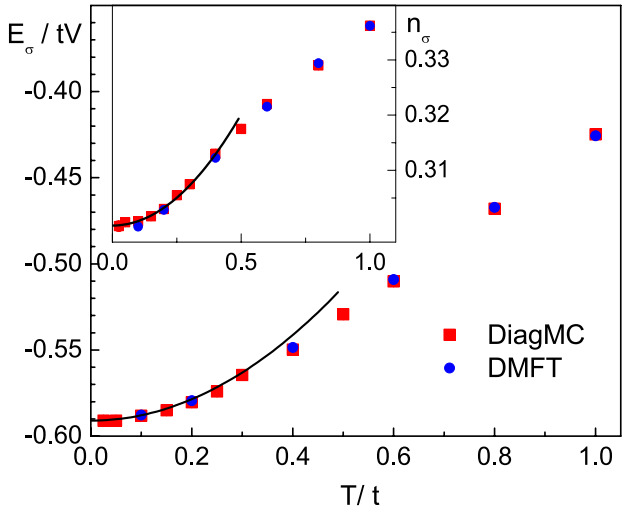}
\caption{ \label{fig:2d_en_a} (color online) Energy density per spin component and density (see inset) as a function of temperature for the 2D Hubbard model at fixed
interaction strength $U/t=4$ and chemical potential $\mu/t=-0.15$.  Within error bars, DMFT (blue circles) coincides with the DiagMC (red squares) results. The Fermi liquid fit, see Eq.(\ref{eq:fit_fermi_liquid}),
is shown by solid lines.
}
\end{center}
\end{figure}

Let us first focus on the two-dimensional heavily hole-doped system with $\langle n_\sigma \rangle \approx 0.3$. Figure \ref{fig:2d_en_a} shows the energy density per spin component $E_{\sigma}/tV$ as a function of temperature at a constant chemical potential $\mu/t=-0.15$ and on-site repulsion $U/t=4$.
The density per spin component $n_{\sigma}$ is shown in the inset.
The lowest temperature in the plot, $\beta t  = 40$, is about two orders of magnitude below the Fermi energy $\epsilon_F$. We clearly observe the Fermi liquid behavior, and the Fermi liquid parameters
can be extracted by fitting the data to the expected the $T^2$ dependence (solid line in the figure)
\begin{equation}
E(T)-E(0) = (\rho_F+\rho_F'\epsilon_F) \frac{\pi^2T^2}{6} \; , \;\;\;
n(T)-n(0) = \rho_F' \frac{\pi^2T^2}{6}  \; ,
\label{eq:fit_fermi_liquid}
\end{equation}
where $\rho_F$ and $\rho_F'$ are the Fermi-surface density of states and its derivative respectively.
We find $\rho_F=0.06(4)$, $\rho_F'=0.095(20)$. Note that in this highly overdoped regime the single-site DMFT results coincide with DiagMC data over the entire temperature range.

\begin{figure}[t!]
\begin{center}
\includegraphics[angle=0, width=0.95\columnwidth]{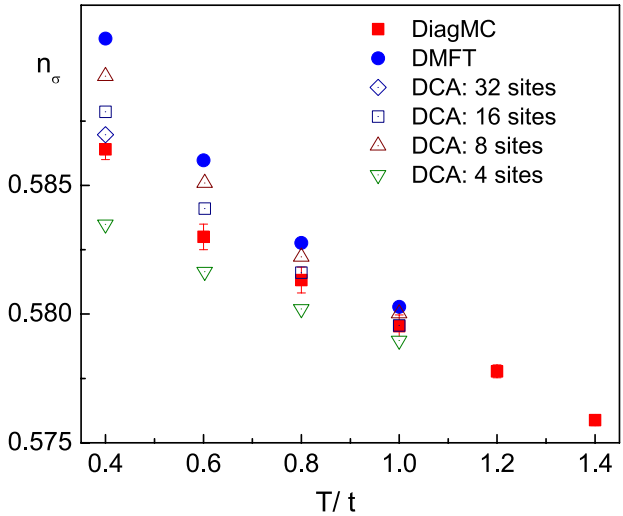}
\caption{ \label{fig:2d_en_b} (color online) Comparison between the DiagMC and DCA methods:
Density as a function of temperature for the 2D Hubbard model at $U/t=4$ and $\mu/t=3.1$.
For temperatures $T \lesssim t$  a systematic deviation of DCA data from DiagMC points is seen.
The deviation is diminishing with increasing the size of the DCA cluster, the four-site results show deviations from single site and larger clusters because of the establishment of local short-range order\cite{plaquette}.
}
\end{center}
\end{figure}

Next, we consider the same system on the particle-doped side at a smaller doping of $20\%$ doping ($\langle n_\sigma\approx 0.6 \rangle$). Figure~\ref{fig:2d_en_b} shows particle density at $\mu/t=3.1$.
In contrast to the previous case, we clearly see a
systematic deviation of the single-site DMFT data from the DiagMC results for temperatures $T \lesssim t$. Upon increasing the cluster size
the DCA data progressively approach the DiagMC points. Still, even $4\times 4$
clusters 
are not sufficient to attain the accuracy of DiagMC in this parameter regime and an extrapolation in the cluster size is needed to reproduce the DiagMC results.

\begin{figure}[t!]
\begin{center}
\includegraphics[angle=0,  width=0.95\columnwidth]{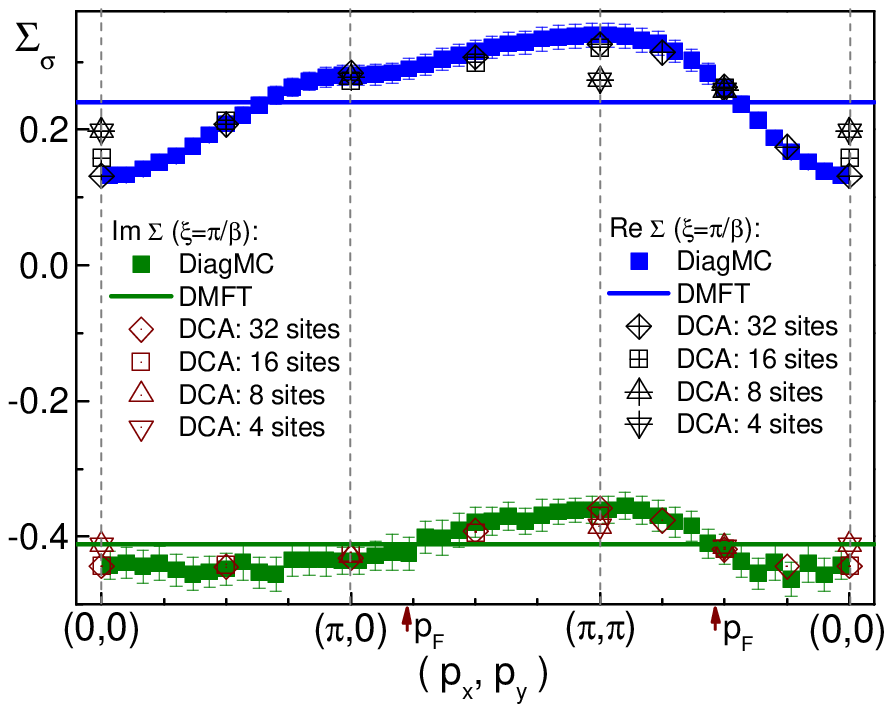}
\caption{ \label{fig:2d_sigma} (color online) Momentum-dependence of the
self-energy at the Matsubara frequency $\xi = \pi/\beta$ along the cut $(0,0)-(\pi,0)-(\pi,\pi)-(0,0)$
in the first Brillouin zone for the Hubbard model with the parameters $U/t=4$,
$\mu/t=3.1$ and $T/t=0.4$.  DiagMC (squares) includes the full
momentum-dependence, whereas single-site DMFT (solid lines) is momentum
independent. The results of DCA calculations are plotted by open symbols for
clusters of size 4, 8, 16, and 32. Note that a good agreement between DiagMC and
DCA is reached with 32-site clusters.  The mean-field contribution (the Hartree
term $ U n_{\sigma} \approx 2.3t$) 
was subtracted to magnify fine details.  The arrows indicate the position of the Fermi momentum $p_F$.
}
\end{center}
\end{figure}

The difference between single-site DMFT  and DiagMC results 
is reflected
in the momentum dependence of the self-energy.
By construction, $\Sigma$ is momentum independent in single-site DMFT.
In Fig.~\ref{fig:2d_sigma}, we consider $\Sigma ({\mathbf p}, \xi)$
at the lowest Matsubara frequency $\xi = \pi/\beta$ along the cut $(0,0)-(\pi,0)-(\pi,\pi)-(0,0)$, in the $(p_x,p_y)$ plane.
Cluster DMFT simulations do include momentum dependence of the self-energy approximately.
As expected, the DCA results get systematically closer to the DiagMC curves when increasing the cluster size. For a cluster of 32 sites, DCA shows a momentum dependence which is in good agreement with DiagMC.

\begin{figure}[t!]
\begin{center}
\includegraphics[angle=0, width=0.95\columnwidth]{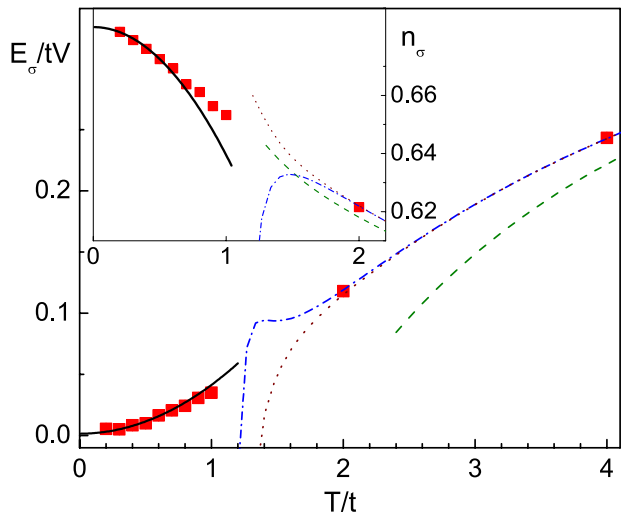}
\caption{ \label{fig:3D_en} (color online) Energy density per spin component $E_{\sigma}/tV$ and density $n_{\sigma}$ (inset) as a function of temperature for the 3D Hubbard model at $U/t=4$ and $\mu' = \mu - U n_{\sigma} = 1.5$. DiagMC results are shown by red squares. The HTSE data are shown by lines: second order in $\beta t$ (green dashed line), eighth order (blue dotted-dashed line), and tenth order (red dotted line). At low temperature, the Fermi liquid behavior is fitted to the DiagMC results (black solid line). The extracted Fermi liquid parameters are $\rho(\epsilon_F)=0.26(4)$, $\rho'(\epsilon_F)=-0.025(2)$.
}
\end{center}
\end{figure}

We now turn to the comparison with high-temperature series expansion
up to tenth order in $\beta t$. To demonstrate that the DiagMC is insensitive to the spatial dimension, we will perform this comparison in three dimensions (3D).
Figure \ref{fig:3D_en} shows the energy and density dependence on temperature for the
3D Hubbard model at $U/t=4$. In this particular example
we keep the renormalized chemical
potential $\mu' = \mu - U n_{\sigma}$ fixed. Within DiagMC, the variable $\mu'$ is more
convenient than the physical chemical potential $\mu$ because the Hartree term  $U n_{\sigma}$
automatically accounts for tadpole diagrams, thus excluding them explicitly from the simulation.
Qualitatively, the data is similar to the 2D case. Without Pad{\'e} approximation the bare HTSE series
give an unbiased answer for temperatures only above $T \sim 2t$, while the degenerate Fermi liquid behavior develops only at $T \lesssim  t$. Only then the energy and density display the characteristic
quadratic dependence on temperature. The fitted values of the density
of single-particle states at the Fermi energy and its derivative are given
in the caption of Fig.~\ref{fig:3D_en}.

The data reported here can serve as established benchmarks for present experiments on ultracold atoms in optical lattices and state-of-the-art numeric techniques. This work constitutes a proof of principle that DiagMC is a reliable method for dealing with hard many-body problems.

To map out the phase diagram of the system one will need to sample diagrams for static response functions and observe the leading instabilities. In the degenerate Fermi liquid regime observed here, the development of these instabilities is controlled by the Fermi liquid theory, which allows one to reliably
predict $T_c$ without actually simulating exponentially low temperatures. Further algorithmic progress can be
made by using Dyson and/or Bethe-Salpeter equations to arrive at the self-consistent skeleton-diagrams description~\cite{Prokofev07}, which would allow one to consider larger interaction strengths and, possibly, reach the critical points of the model. Going below the critical point  might be possible by introducing anomalous propagators.

\begin{acknowledgments}
We thank V. Scarola and J. Oitmaa for providing us with the high-temperature expansion series. We are grateful to A.~Georges,  A. Muramatsu, O. Parcollet, F. Werner and P. Werner for stimulating discussions.
We wish to acknowledge support from FWO-Vlaanderen, the Swiss National Science Foundation, NSF Grants PHY-0653183 and DMR-0705847,  the DARPA OLE program, and the Aspen Center for Physics.  Simulations were performed on the Brutus cluster at ETH Zurich.
The DMFT results were obtained with the numerically exact continuous-time auxiliary field cluster impurity solver\cite{ct-aux}, based on the ALPS library \cite{ALPS}.
\end{acknowledgments}


\begin{thebibliography}{99}
\bibitem{Troyer} M. Troyer and U.-J. Wiese, Phys. Rev. Lett. \textbf{94}, 170201 (2005).
\bibitem{Hubbard63} J. Hubbard, Proc. Roy. Soc. (London), Ser. A {\bf 276}, 238 (1963).
\bibitem{Anderson} P.W. Anderson, {\it{The Theory of Superconductivity in High-$T_c$ Curpates}} (Princeton Univ. Press, Princeton, 1997).
\bibitem{Esslinger05} M. K{\"o}hl {\it et al.}, Phys. Rev. Lett. {\bf 94}, 080403 (2005).
\bibitem{Jordens08}  R. J\"ordens, N. Strohmaier, K. G\"unter, H. Moritz  and  T. Esslinger, Nature {\bf 455}, 204 (2008).
\bibitem{Schneider08} U. Schneider, L. Hackermuller, S. Will, Th. Best, I. Bloch, T.A. Costi, R.W. Helmes, D. Rasch, A. Rosch, Science {\bf 322}, 1520 (2008).
\bibitem{Trotzky09} S. Trotzky {\it et al.}, cond-mat/0905.4882 (2009).

\bibitem{Prokofev98} N.V. Prokof'ev and B.V. Svistunov, Phys. Rev. Lett. {\bf 81}, 2514 (1998).

\bibitem{Fetter} A.L. Fetter and J.D. Walecka, {\it Quantum Theory of Many-Particle Systems} (McGraw-Hill, New York, 1971).
\bibitem{Landau} E.M. Lifshitz and L.P. Pitaevskii, \textit{Statistical Mechanics}, Part 2  (Pergamon Press, Oxford, 1980).	

\bibitem{Prokofev08} N.V. ProkofÕev and B.V. Svistunov, Phys. Rev. B {\bf 77},  020408 (2008); {\it ibid.} 125101 (2008).
\bibitem{Dyson} F.J. Dyson, Phys. Rev. {\bf 85}, 631 (1952).
\bibitem{Henderson92} J.A. Henderson, J. Oitmaa and M.C.B. Ashley, Phys. Rev. B {\bf 46}, 6328 (1992).
\bibitem{VanHoucke08} K. {Van Houcke}, E. Kozik, N.V. Prokof'ev, and B.V. Svistunov, in Computer Simulation Studies in Condensed Matter Physics XXI, edited by D.P. Landau, S.P. Lewis, and
H. B. Sch\"uttler (Springer-Verlag, Berlin, 2008); arXiv:0802.2923.
\bibitem{Georges96} A. Georges, G. Kotliar, W. Krauth, and M.J. Rozenberg,  Rev. Mod. Phys. {\bf 68}, 13  (1996).
\bibitem{Hettler98}M.~Hettler, A. N. Tahvildar-Zadeh, M. Jarrell, {\it et al.}, Phys. Rev. B {\bf 58}, R7475 (1998).
\bibitem{Lichtenstein00} A. I. Lichtenstein and M. I. Katsnelson, Phys. Rev. B {\bf 62}, R9283 (2000).
\bibitem{Kotliar01}G. Kotliar, S. Savrasov, G. P\'alsson, and G. Biroli, Phys. Rev. B {\bf 58}, R7475 (1998).
\bibitem{Maier05} T. Maier, M. Jarrell, T. Pruschke, and M.H. Hettler,  Rev. Mod. Phys. {\bf 77}, 1027 (2005).
\bibitem{tenHaaf92} D.F. ten Haaf and J. M. van Leeuwen, Phys. Rev. B {\bf 46}, 6313 (1992).


\bibitem{SCET} C.W. Bauer, S. Fleming, D. Pirjol, I.W. Stewart, Phys.Rev. D \textbf{63} 114020 (2001).

\bibitem{D-Sch} R. Alkofer, L. von Smekal, Phys.Rept. \textbf{353} 281 (2001).

\bibitem{NPT} F. Di Renzo, M. Laine, Y. Schroder, C. Torrero, JHEP \textbf{09}, 061 (2008).

\bibitem{Prokofev07} N.V. ProkofÕev and B.V. Svistunov, Phys. Rev. Lett. {\bf 99}, 250201 (2007).





\bibitem{plaquette}E. Gull, P. Werner, X. Wang, {\it et al.}, EPL {\bf 84}, 37009 (2008).
\bibitem{ct-aux}E. Gull, P. Werner, O. Parcollet, and M. Troyer, EPL {\bf 82},57003 (2008).
\bibitem{ALPS}F. Albuquerque {\it et al.}, J. Magn. Magn. Mater. {\bf 310}, 1187 (2007).
\end{thebibliography}
\end{document}